\documentclass[floatfix]{article}
\usepackage{aaai}
\usepackage{graphics}
\title{Social Media as Windows on the Social Life of the Mind}
\author{Cosma Rohilla Shalizi\\ Statistics Department, Carnegie Mellon University\\ 5000 Forbes Avenue\\ Pittsburgh, PA 15213 USA}
\begin{document}
\maketitle

\begin{abstract}
  This is a programmatic paper, marking out two directions in which the study
  of social media can contribute to broader problems of social science:
  understanding cultural evolution and understanding collective cognition.
  Under the first heading, I discuss some difficulties with the usual,
  adaptationist explanations of cultural phenomena, alternative explanations
  involving network diffusion effects, and some ways these could be tested
  using social-media data.  Under the second I describe some of the ways in
  which social media could be used to study how the social organization of an
  epistemic community supports its collective cognitive performance.
\end{abstract}

Let me begin by considering two\footnote{Of course, people think a lot about
  their own and others' social interactions, and a big use of social media is
  sharing these thoughts.  But in this social media are no different from any
  other form of human, or for that matter primate, association.} senses in
which we might speak of human thought as being ``social'', and how they might
orient the study of social information processing and social media.

The first sense is a common-place of many schools in the social sciences and
humanities: our thought relies on the cultural transmission of cognitive tools.
Every individual thinker, no matter how innovative or even lonely they may be,
depends crucially on a vast array of cognitive tools (concepts, procedures,
languages, assumptions, values, ...) which they did not devise themselves, and
could not have devised for themselves.  Instead they inherited these cognitive
tools from interacting with other people, who for the most part themselves did
not invent them.
\cite{Dewey-public,Vygotsky-thought-and-language,Popper-open-society,%
  Balkin-cultural-software}\footnote{``[K]nowledge is a function of association
  and communication; it depends upon tradition, upon tools and methods socially
  transmitted, developed and sanctioned.  Faculties of effectual observation,
  reflection and desire are habits acquired under the influence of the culture
  and institutions of society, not ready-made inherent powers''
  \cite[p. 158]{Dewey-public}.  Cf.\ \cite[ch.\ 23--24]{Popper-open-society}.}
(Whether this dependence on tradition is a logical necessity, or merely a
reflection of our peculiar bounded rationality and bounded lifespan, is a deep
question, fortunately not relevant here.)  While individual thinkers invent and
discover, it is nonetheless true that innovations are typically refined,
extended and perfected by groups, and that it is very rare indeed for highly
developed concepts and ideas to emerge from a single, isolated thinker, rather
than from a process of interaction
\cite{Toulmin-human-understanding,Kitcher-advancement,Collins-philosophies,%
  Ziman-real-science}.

The branches of social science for which these facts are common-places have
largely developed them philosophically
\cite{Toulmin-human-understanding,Turner-brains-practices}, or qualitatively
\cite{Vygotsky-thought-and-language,Vygotsky-mind-in-society,%
  Balkin-cultural-software,Mercer-words-and-minds} or even ethnographically
\cite{Luria-cognitive-development,Hutchins-cognition-in-the-wild}.  (But see
\cite{Elements-of-reason}.)  In part this has been for reasons of cultural and
intellectual politics, as the relevant scholars have tended to fall on the
``interpretation'' rather than ``explanation'' side of the divide in the social
sciences \cite{Sperber-explaining}, so that attention to the social nature of
thought often goes along with more or less pronounced hostility to quantitative
and computational modeling
(e.g. \citeauthor{Hutchins-cognition-in-the-wild,Mercer-words-and-minds}).
This supposed opposition is thoroughly mis-guided \cite{Frawley-on-Vygotsky},
but it is not likely that anyone will be argued out of it any time soon.

More promisingly, however, one {\em good} reason for developing this idea
through small-scale qualitative studies has been that it was impossible to
gather {\em relevant} data, suitable for quantitative analysis, on any large
scale.  With the rise of social media, however, many people are, for their own
purposes, generating exactly this kind of data for us --- traces of their
communicative interactions as they work out their thoughts about matters of
common concern.  They are doing so on a wide range of subjects, under a wide
range of different institutional mechanisms which structure their interactions
in many different ways, creating natural sources of variation which the social
scientist can try to exploit to learn more about the effects of subject matter,
of communicative structure, and of other factors on cultural dynamics, and
perhaps ultimately even on innovation and discovery.  The next section points
out some of the outstanding problems and methodological pitfalls of this area.

The {\em other} important sense in which human thought can be ``social'' is
that it seems to make sense to regard at least some human social institutions
as, themselves, information-processing systems, engaging in computations which
cannot be localized to representations in the mind of any one of their members.
On large scales, market economies, corporations and other bureaucracies,
scientific disciplines, and democratic polities all have something of this
collective information-processing character.  Knowing how they accomplish this
would be deeply rewarding, and, if that understanding can be used to make them
work better, of profound economic and political importance.  A frontal assault
on this problem, as represented by one of those grand institutions, is unlikely
to succeed (though it may be a magnificent failure).  Fortunately, social
information processing also occurs in much humbler institutions, such as
tagging systems and collaborative filtering, where issues of data collection
and even experimental manipulation are much more manageable, and where we might
hope to learn more, before tackling the fundamental problems of social
science. I will lay out some of what should be on the agenda of the study of
social information processing, in particular points of contact with machine
learning.

\section{Cultural Evolution}
\label{sec:cult-ev}

``Culture is the precipitate of cognition and communication in a human
population'' \cite{Sperber-explaining}.  That is, cultural traits --- beliefs,
practices, habits, conventions, expressions, norms --- are not just ones which
are common across a population, but ones which are spread across a population
because its members communicate with one another.  (Knowing that it's painful
to look at the sun directly is not cultural; knowing that the direction in
which the sun rises is called ``east'' is cultural.)  Cultural phenomena are
thus emergent, the result of the communicative interaction of cognitive agents.
If we are to understand how cultures work, we need to understand something
about both parts, the internal cognitive mechanisms and the effects of
different patterns of interaction.  Social media offer a window into the
communicative part of the problem of unrivaled clarity and breadth.  This is
extremely exciting, but in looking through this window we should bear in mind
some methodological difficulties to interpreting the view through this window.

It is a common-place observation that there are strong relationships between
cultural traits and social attributes; that different social groups accept and
transmit different bits of culture.  Most attempts to explain this from within
the social sciences (emphatically including historical materialism
\cite{Elster-on-Marx,Cohen-on-Marx-on-history} and its variants) argue that
this is due to some causal influence of social organization on the {\em
  content} of culture.  (``Social being determines consciousness''
\cite{Marx-Engels-german-ideology} --- or, once the Hegelian gas has been
released, social life shapes thought.)  In these views, culture varies with
social position {\em because} the former is adapted to the latter, or reflects
it, or expresses it.

It is natural for us, as beings acutely sensitive to nuances of cultural
meanings, to try to explain cultural differences by trying to explain the {\em
  content} of widely-shared, cultural representations.  It is natural to
suppose that, say, one news story rises to the top of a social aggregation
system because it is {\em more interesting} than other stories which did not.
Such explanations are even valid a lot of the time.  It is nonetheless
important, as a point of methodological hygiene, to develop ways of telling
when some bit of culture succeeds in propagating because its content fits its
circumstances, if only because, being creatures acutely sensitive to nuances of
cultural meanings, it is far too easy for us to spin such stories no matter
what the truth might be.  \citeauthor{Lieberson-matter-of-taste}
\shortcite{Lieberson-matter-of-taste} points out that many widely-accepted
explanations of trends in fashions, children's names, etc., cannot possibly be
right (because, e.g., the trend pre-dates or is more widely spread than the
supposed cause), and that these are instead better explained by purely internal
mechanisms of the respective fields.  In biology, adaptive and non-adaptive
evolution are demarcated by means of {\em neutral models}.  These are models of
the genetic changes which would be expected due to reproductive mechanisms and
chance alone, all genetic variants being assumed to be ``adaptively neutral'',
i.e., of equal fitness.  Only when actual populations depart markedly from the
predictions of neutral models can adaptation be (reliably) inferred
\cite{Nitecki-Hoffman-neutral-models,Harvey-Pagel}.  Before the student of
social media, or other cultural media, can start explaining phenomena by
reference to content, they need to check that there actually is something to be
explained.

A highly simplistic model may make this point more concrete.  Consider a
network in which people have two binary traits, one of which is stable (we may
think of this as ``class'' or ``race'' or some similar status), and the other
is changeable (think of fashions, or political opinions).  Assume that the
network is assortative on the stable, social-type trait, so that people are
more likely to be linked to others of the same type than those of a different
type.  Such ``assortativity'' or ``homophily'' is observed in many, perhaps
most social networks, often on such stable social-status type variables
\cite{Birds-of-a-Feather-review,MEJN-mixing-patterns}.  Now assign the cultural
trait to people uniformly and independently of their social trait (or anything
else).  Initially, then, there will be no correlation between social and
cultural traits, and no assortativity based on culture.

We might expect such correlations to appear if the process of cultural
transmission and retention is biased --- if, say, certain cultural values only
make sense for those in certain social positions.  In that case, we would
expect to find a growing ``fit'' between cultural and social variables, as the
former adapt to the latter.  But by this point you will not be surprised to
learn that {\em neutral} transmission processes can also induce such
correlations.  To be specific, let's implement the ``voter model''
\cite{Liggett-particle-systems}: at each discrete time step, a node is chosen
uniformly at random, independently of past and future choices.  This node
chooses a neighbor (again uniformly and independently), and copies its value of
the cultural trait.

\begin{figure}[tp]
\begin{center}
\resizebox{\columnwidth}{!}{\includegraphics{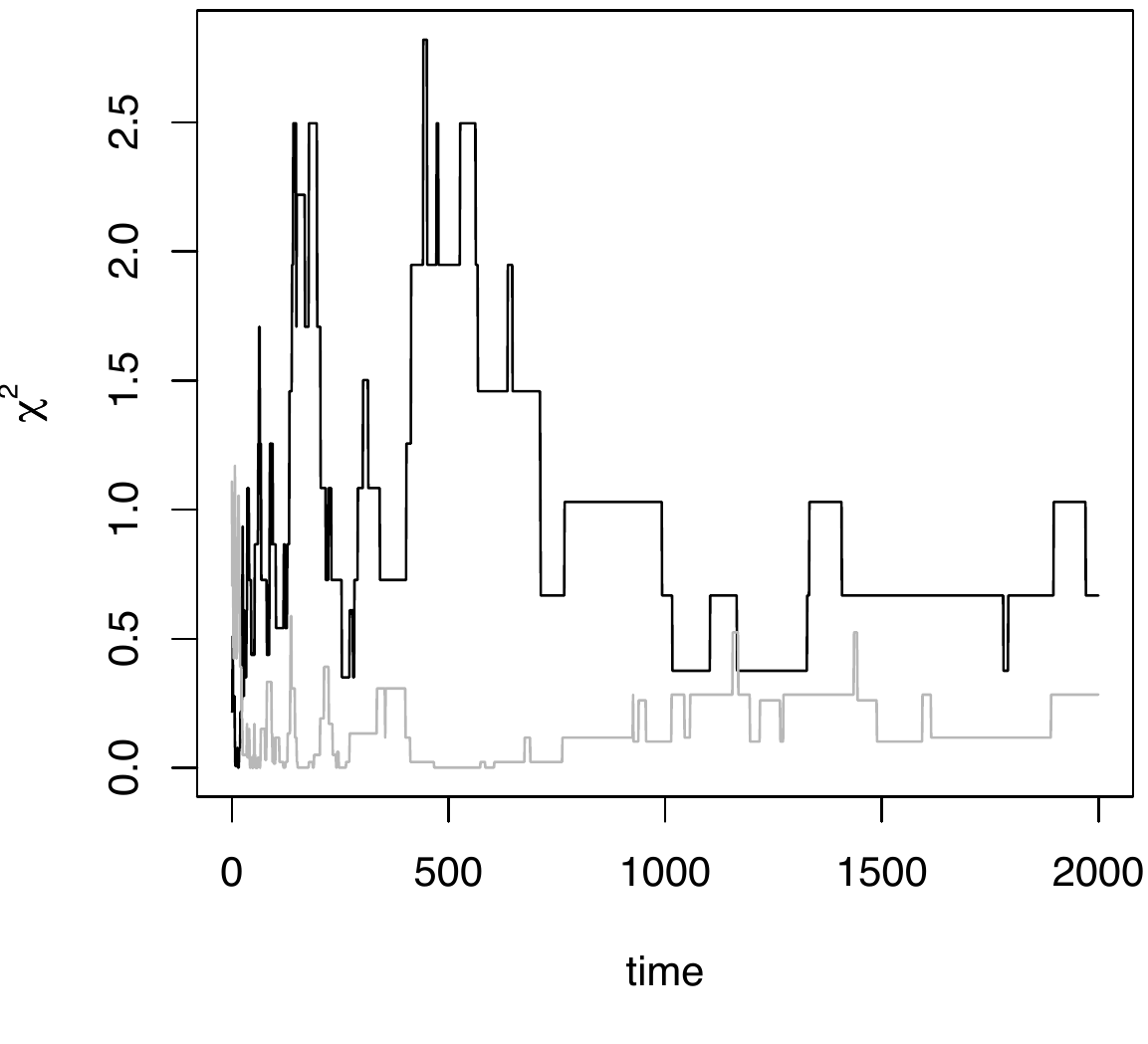}}
\end{center}
\caption{{\small Neutral copying induces correlations between social and
    cultural traits in assortative networks.  The graph has 100 nodes, randomly
    divided between two social types (equally probable), and a binary-valued
    cultural trait (initially equally probable).  Edges between nodes of the
    same type occur with probability $p_1$, those between different types have
    probability $p_2$.  At each time step, a random node copies the cultural
    trait of a random neighbor.  Horizontal axis: time.  Vertical axis:
    $\chi^2$ statistic for the correlation between the social and cultural
    variables.  Black line: behavior of an assortative network ($p_1 = 0.09$,
    $p_2 = 0.01$, assortativity coefficient \cite{MEJN-mixing-patterns} of
    realized graph $r = 0.80$).  Note the eventual decline of $\chi^2$ as the
    network moves towards a homogeneous equilibrium; in the very long run it
    will reach 0.  Grey line: behavior of a non-assortative network ($p_1 = p_2
    = 0.05$, $r = 0.045$).}}
\label{fig:chisq-time-series}
\end{figure}

Clearly, in a connected network, there are two absorbing states, which are
culturally homogeneous, and eventually the network must settle into one or the
other of them, but the time it takes to do so will typically be quite long
\cite{Sood-Redner-voter-model}.  In the meanwhile, if the network is socially
assortative, numerical experiments (Fig.\ \ref{fig:chisq-time-series}) show
that the social and the cultural traits tend to become correlated during a long
``meta-stable'' period.\footnote{One could say that the cultural trait of a
  node is still ``in the final analysis'' determined by its social type, but
  only with the proviso that the over-all structure of the network ``screens
  off'' the latter, rendering it causally irrelevant
  \cite{Galles-Pearl-axioms}.}  If I'd said that the social types were ``lower
class'' and ``middle class'', and the cultural traits ``likes black velvet
paintings'' and ``likes black and white photographs'', the temptation to
explain the correlation by content would be overwhelming
\cite{Bourdieu-distinction}.  Nonetheless, which way the correlation went would
be a matter of pure chance, or more exactly of the reinforcement and
amplification of small fluctuations, though {\em some} such pattern forms with
high probability.  (This contrast between long- and medium- run behavior is not
uncommon in self-reinforcing network processes
\cite{Pemantle-Skyrms-network-formation}.)  The strength of the
dynamically-induced correlations depends on the assortativity of the social
network; if it is not assortative, then the correlations between social and
cultural traits only rarely rise above the levels to be expected by chance
(Fig.\ \ref{fig:chisq-time-series}).

Social scientists interested in communications have appreciated for a long time
that network structure is very important to {\em how} information flows through
a social group \cite{Katz-Lazarsfeld-personal-influence,%
  Huckfeldt-Johnson-Sprague-political-disagreement}, but they have not, so far
as I know, realized that it can create just the kind of correlation that seems
to cry out for an explanation by content.  In fact, the real situation is
somewhat worse than this, because it really isn't a given that people change
because of interacting with their neighbors.  It could well be that people have
neighbors who are similar to themselves, and so they all respond similarly to
common exogenous causes, without any direct interactions
\cite{Steglich-Snijders-Pearson-selection-and-influence}.\footnote{This
  possibility seems to confound the claims of the recent, and
  widely-publicized, study of the spread of obesity in a social network
  \cite{Christakis-Fowler-spread-of-obesity}.}  If one thinks of trying to
explain why certain users prefer certain kinds of news stories, for example,
one must account not only assortativity, but also for common exposure to some
outside news source.  --- None of this, incidentally, requires that people {\em
  actually} make decisions randomly, but only that the reasons which lead them
to their decisions are effectively unpredictable from the other variables in
the system.

The moral is not that these kinds effect explain {\em all} correlations between
social and cultural traits, or even between different cultural traits.  Rather,
it shows that a neutral explanation is logically possible.  To support an
adaptive explanation of a correlation, then, one must show some way in which
the neutral model is not adequate to the data.  For example, additional
experiments (not shown) indicate that, if I take the model simulated in Fig.\
\ref{fig:chisq-time-series} and break the graph into communities (following
\citeauthor{MEJN-Girvan-community-structure}
\shortcite{MEJN-Girvan-community-structure}), then social type and cultural
traits are {\em conditionally} independent, given community membership, even in
strongly-assortative networks.  This conditional independence does {\em not}
hold when different social types have differing biases for or against various
cultural traits.  Only when we have found and verified such discrepancies
between our data and the predictions of a good neutral model can we say that
the adaptive explanation has passed a severe test and truly has evidence in its
support \cite{Mayo-error}.

\section{Collective Cognition}
\label{sec:coll-cog}

It's been recognized since the 1930s that market economies are ``collective
calculating devices''
\cite{Lange-Taylor-socialism,Hayek-individualism-and-economic-order}.  A
market-clearing allocation of good and services is simply too big for anyone to
grasp, let alone find.  Instead it is the process of exchange itself which
adaptively finds and implements this allocation.\footnote{On the formal
  computational power of market-like systems, see
  \cite{Walsh-et-al-markets-and-satisfiability}.}  This is an example of what
we might call {\em collective cognition}, by analogy to the classical
\cite{Olson-logic-of-collective-action} ``collective action''.  Similarly, the
problems of designing policies for governments are largely beyond the scope of
what anyone can actually do, but not beyond the scope of democratic
deliberation, which reduces the problem from solving for the optimal policy in
one stroke, to criticizing and improving policies piecemeal
\cite{Braybrooke-Lindblom}, in light of the information and ideas of many
participants.  \cite{Popper-open-society,Lindblom-intelligence,%
  Ober-athenian-legacies} (Historically, democratic decision-making has been
associated with more social power than other forms of government
\cite{McNeill-pursuit-of-power}, but the causality is unclear.)  Similar
remarks apply to bureaucratic organizations, such as corporations, and to
scientific disciplines.

It is notable that modern societies are vastly better at collective cognition
than earlier ones.  The degree of organization, and its precision, which we
take for granted would have been astonishing for even the inhabitants of the
most advanced societies c. 1600, to say nothing of c. 100.  Historians have
explored some of the technical and institutional underpinnings of these
organizational revolutions \cite{McNeill-pursuit-of-power,Beniger-control,%
  Yates-control}, but at a deeper level we have little idea {\em why} this is
so, or why what we do works (when it does work).  This makes it harder to {\em
  improve} the functioning of our institutions for collective cognition.
Economic theories of mechanism design attempt to do so, but largely address the
problem of {\em motivating} people to act in certain ways, rather than of how
to figure out what the right action is \cite{Miller-managerial-dilemmas}.

These are all very large themes indeed, of course, and it might seem grandiose
to even mention them in this context.  I am not suggesting that studying social
media will give us the key to all organization technologies.  What it can do,
however, is give us a set of case studies where, on a much humbler level,
people are nonetheless engaged in social information processing and collective
cognition.  Just as no one market participant decides on or represents the
over-all market allocation, and no one scholar ever grasps more than a small
portion of what is known about conic sections or cellular slime molds, the
movies or bookmarks which get recommended by collaborative filtering services
are the emergent products of the interactions of many participants
\cite{Lerman-social-info-proc-in-social-news-agg}.  What social media offer us,
again, is the possibility to automatically collect large-scale data on such
phenomena, combined with a clear understanding of the interaction structure (or
at least a lot of it), as well as much of the external circumstances and the
goals of the group.  We can thus begin, at least at a small scale, to begin
building and systematically testing theories which explain how social
information processing and collective cognition succeed when they do.

It might be thought that the theoretical explanation is rather simple, and goes
(currently) under the name of ``the wisdom of crowds''
\cite{Surowiecki-crowds}: individuals make noisy guesses, which on average are
unbiased and uncorrelated, so simple averaging leads to convergence on the
appropriate answer.  Taken seriously, this explanation implies that our
economy, our sciences and our polities manage to work {\em despite} their
social organization, that science (for example) would progress much faster if
scientists did not collaborate, did not read each others' papers, etc.  While
every scientist feels this way occasionally, it is hard to take seriously.
Clearly, there has to be an explanation for the success of social information
processing {\em other than} averaging uncorrelated guesses, something which can
handle, and perhaps even exploit, statistical dependence between decision
makers.

A particularly interesting line of attack on these problems is suggested by the
analogy with ensemble methods in machine learning.  As
\citeauthor{Domingos-on-Occams-Razor} \shortcite{Domingos-on-Occams-Razor} has
pointed out, the success of these methods seems to confound naive
interpretations of Occam's Razor, in much the same way that the success of
social information processing confounds the simple ``wisdom of the crowds''
story.  Ensemble methods, in which large numbers of low-capacity classifiers or
predictors (e.g., shallow classification trees) are combined, effectively
create a single model of what appears to be very high capacity, and so they
appear to be nothing but an invitation to over-fitting.  Worse, typically
ensemble methods such as boosting \cite{tEoSL}, bagging \cite{Breiman-bagging}
and mixtures of experts \cite{Jacobs-bias-variance-of-mixtures-of-experts}
create {\em correlated} low-level predictors, so that the simple
average-the-crowd story is inapplicable.  In fact, it is precisely because the
component predictors are correlated, but not identical, that the {\em actual}
capacity of the ensemble is much smaller than its {\em apparent} capacity.

A similar result holds for cooperative problem-solving
\cite{Hong-Page-diversity-rules}.  Under mild conditions, it can be shown that
a large group of ``weak'' heuristic problem-solvers, whose performance in
isolation is only slightly better than random search, will actually out-perform
a similarly-sized group of ``strong'' heuristics, ones whose average
performance in isolation is much better.  One of those conditions, however, is
that the problem-solvers must be able to communicate with each other, making
their candidate solutions strongly dependent rather than uncorrelated.  There
is good evidence that this beneficial effect of heuristic diversity and
communication is actually seen in the cognitive performance of human groups
\cite{scotte-Difference}.  This suggests a very promising direction for
research on social information processing, namely to use the mathematical
techniques of statistical learning theory to establish bounds on the
performance of suitable sorts of ensemble-learners and group problem-solvers,
and see how close actual social information processing systems come to attain
those bounds, and how the latter could be improved by changes to their
architectures.

Both ensemble methods and the \citeauthor{Hong-Page-diversity-rules} results on
diverse heuristics posit relatively simple forms of ``social'' organization,
such as direct averaging, or passing a problem to the next person able to
improve on the current solution.  There is every reason to think, however, that
the optimal form of organization will actually depend on the structure of the
problem being solved.  (Cf. \citeauthor{Braybrooke-Lindblom}
\shortcite{Braybrooke-Lindblom} on how the social organization of policy
analysts serves their cognitive strategy of ``disjointed incrementalism.'')  In
particular, coordination over time is not an issue in ensemble methods, and
handled by assumption in the \citeauthor{Hong-Page-diversity-rules} model, but
extremely important in real-world systems for social information processing and
collective cognition.

This suggests a final line of research, one which draws together ideas from
distributed systems, economics and statistical mechanics.  Experience with
distributed systems shows that often the hardest part of their design is
ensuring coordination over time, and that failure to do so can lead to all
manner of unwanted behavior, in particular to wild oscillations and/or locking
into deeply undesirable configurations \cite{Lynch-on-distributed-algorithms}.
In fact, the failure modes of distributed systems are strongly reminiscent of
the pathologies of economic \cite{Chamley-herds} and statistical-mechanical
\cite{Young-strategy-structure} models of social learning, when they are placed
in suitable (that is, unsuitable) situations.  Designing, or reforming, a
system for computer-mediate social information processing is at once a problem
of distributed algorithm design {\em and} a problem of mechanism design, and
they two modes or aspects should inform one another, as well as empirical
results about what actually happens when real human beings use different
systems for different tasks.

\paragraph*{Acknowledgments} Thanks to P. Agre, P. Domingos, C. Genovese,
K. Lee, S. Page, W. Tozier, E. Smith, N. Snoad, and the participants of the
2002 workshop on collective cognition and distributed intelligence at the Santa
Fe Institute for valuable discussions, and to K. Klinkner for many reasons
(including valuable discussions).

\bibliography{locusts}
\bibliographystyle{aaai}

\end{document}